\documentclass[12pt]{article}
\usepackage{graphicx}
\begin{document}
\begin{center}
{\large\bf Spectrum of Cosmic Microwave Fluctuations and the
Formation of Galaxies in a Modified Gravity Theory} \vskip 0.3
true in {\large J. W. Moffat} \vskip 0.3 true in {\it The
Perimeter Institute for Theoretical Physics, Waterloo, Ontario,
N2L 2Y5, Canada} \vskip 0.3 true in and \vskip 0.3 true in {\it
Department of Physics, University of Waterloo, Waterloo, Ontario
N2L 3G1, Canada}
\end{center}
%\date{\today}
\begin{abstract}%
A modified gravity (MOG) possesses a light, neutral vector
particle called a ``phion'' associated with a vector field
$\phi^\mu$, which forms a cold fluid of Bose-Einstein condensates
before recombination with zero pressure and zero shear viscosity.
The energy density associated with this Bose-Einstein condensate
fluid dominates the energy density before recombination and
produces a density parameter, $\Omega_\phi\sim 0.3$, that together
with the fractional baryon density $\Omega_b\sim 0.04$, and a
cosmological constant parameter $\Omega_\Lambda\sim 0.7$ yields an
approximate fit to the data for the acoustical oscillations in the
CMB power spectrum. The quantum phion condensate fluid is abundant
well before recombination and can clump and form the primordial
structure for galaxies. At late times in the expanding universe,
in local bound systems such as galaxies ordinary baryonic matter
dominates the matter density. For galactic systems in the present
epoch, the modified Newtonian acceleration law determined by MOG
describes well galaxy rotation curve data and X-ray cluster mass
profile data.
\end{abstract}
\vskip 0.2 true in e-mail: john.moffat@utoronto.ca

%\pacs{ }

\section{Introduction}

A relativistic modified gravity (MOG) theory~\cite{Moffat,Moffat2}
has been proposed to explain the rotational velocity curves of
galaxies and the X-ray data for clusters of galaxies with a
modified Newtonian acceleration law, without non-baryonic dark
matter. A fitting routine for galaxy rotation curves has been used
to fit a large number of galaxy rotational velocity curve data,
including low surface brightness (LSB), high surface brightness
(HSB), dwarf galaxies and elliptical galaxies with both
photometric data and a two-parameter core model without
non-baryonic dark matter~\cite{Moffat2,Brownstein}. The fits to
the data are remarkably good and for the photometric data only the
one parameter, the mass-to-light ratio $\langle M/L\rangle$, is
used for the fitting, once two parameters $M_0$ and $r_0$ are
universally fixed for galaxies and dwarf galaxies. The fitting
results for galaxies are consistent with the data for the
Tully-Fisher relation~\cite{Tully}. A large sample of X-ray mass
profile cluster data has also been fitted~\cite{Brownstein2}. The
gravity theory also fits the anomalous acceleration data observed
during the Pioneer 10-11 spacecraft missions~\cite{Moffat3}, and
is consistent with gravity observations in the solar system and
the binary pulsar PSR 1913+16.

The MOG requires that Newton's constant G, the coupling constant
$\omega$ that measures the strength of the coupling of a skew
field to matter and the mass $\mu$ of the skew field, vary with
distance and time, so that agreement with the solar system and the
binary pulsar PSR 1913+16 data can be achieved, as well as fits to
galaxy rotation curve data and galaxy cluster data. The variation
of $G$ leads to a consistent description of relativistic lensing
effects for galaxies without non-baryonic dark matter. In
ref.~\cite{Moffat2,Reuter}, the variation of these constants was
based on a renormalization group (RG) flow description of quantum
gravity theory formulated in terms of an effective classical
action. Large infrared renormalization effects can cause the
effective $G$, $\omega$, $\mu$ and the cosmological constant
$\Lambda$ to run with momentum $k$ and a cutoff procedure leads to
a space and time varying $G$, $\omega$ and $\mu$, where
$\mu=1/r_0$ and $r_0$ is the effective range of the skew symmetric
field. In the MOG theory~\cite{Moffat2}, the action contains a
contribution that leads to {\it effective} field equations that
describe the variations of $G$, $\omega$ and $\mu$. In principle,
we can solve for the complete set of field equations and determine
the dynamical behavior of all the fields. However, in practice we
make approximations allowing us to obtain partial solutions to the
equations, yielding predictions for the various physical systems
considered.

A modified gravity theory must be able to explain the following
cosmological data and phenomena:
\begin{enumerate}

\item The cosmic microwave background (CMB) data, including the power
spectrum for both large and small angular scales on the sky;

\item The formation of galaxies before and after recombination;

\item Gravitational lensing;

\item N-body simulations of large-scale galaxy surveys;

\item The accelerating expansion of the universe.

\end{enumerate}

The standard explanation for these phenomena is the cold dark
matter (CDM) scenario and some form of uniform dark energy that
begins to dominate in the present epoch of the universe. The
problem with this scenario is that 96\% of the universe's matter
and energy budget is invisible. It remains undetected in
ground-based laboratory experiments, and has yet to be given a
satisfactory physical interpretation~\cite{Baudis}. An attempt to
explain the CMB power spectrum at the surface of last scattering
about 370,000 years after the big bang in terms of the visible
baryon number density~\cite{McGaugh}, leads to a satisfactory fit
to the WMAP and Archeops data~\cite{WMAP,Archeops,Boomerang} for
the first two peaks in the power spectrum at small angles with the
baryon density fraction $\Omega_b\sim 0.04$ and the cosmological
constant density fraction $\Omega_\Lambda\sim 0.95$. This scenario
requires a neutrino mass contribution with $m_\nu\sim 1-2$ eV. The
baryon density parameter $\Omega_b\sim 0.04$ is in agreement with
recent estimates for this quantity at big bang nucleosynthesis.
The problem with this pure baryon-neutrino scenario is that a
third putative peak in the angular power spectrum is erased by
baryon drag. A problem with neutrinos describing dark matter is
that they become non-relativistic at a late-time epoch in the
expanding universe, and their fluctuations are erased in the power
spectrum.

The most recent balloon-born Boomerang data~\cite{Boomerang}
indicate the existence of a third peak in the power spectrum,
which is erased by baryon drag without cold dark
matter~\cite{Silk}. If future data confirms that a significant
third peak in the angular power spectrum exists, then this
strongly supports that some form of dark, non-baryonic matter
should exist at the surface of last scattering. All of this
suggests that some kind of non-relativistic dark matter should be
accounted for in cosmology. On the other hand, the MOG fits to the
galaxy rotation curve data and the X-ray galaxy cluster
data~\cite{Brownstein,Brownstein2} without non-baryonic dark
matter are remarkably good and avoid problematic issues related to
CDM fits to the galaxy data.

It is also necessary to provide a satisfactory explanation for the
early universe growth of structure and the eventual emergence of
galaxies and clusters of galaxies. Large computer N-body
simulations show that some kind of non-relativistic dark matter
has to constitute the proto-galaxies about 200 million years after
the big bang~\cite{Moore,White}. The data suggest that early
universe density fluctuations were a Gaussian random field
described by a cold dark matter (CDM). The standard assumption is
that the CDM is made of elementary particles that interact only
gravitationally. Baryons and neutrinos alone fail to produce a
realistic description of large-scale galaxies and clusters
surveys.

An important extra-degree of freedom in MOG is a light,
electrically uncharged vector particle called a ``phion''. In the
early universe at a temperature $T < T_c$, where $T_c$ is a
critical temperature, the phions become a Bose-Einstein condensate
(BEC) fluid. The phion condensates couple weakly with
gravitational strength to ordinary baryonic matter. This cold
fluid of phion condensates dominates the density of matter at
cosmological scales and, because of its clumping due to
gravitational collapse, allows the formation of structure and
galaxies at sub-horizon scales well before recombination. We do
not postulate the existence of cold dark matter in the form of
heavy, {\it new stable particles} such as supersymmetric WIMPS.

In the following, we shall calculate the power spectrum for values
$l \gg 1$ obtaining the positions and heights of the acoustical
peaks from cosmological parameters. We shall use the analytical
formula derived by Mukhanov~\cite{Mukhanov} and earlier work by Hu
and Sugijama~\cite{Hu}, Frampton, Ng and Rohm ~\cite{Frampton},
Weinberg~\cite{Weinberg} and Dodelson~\cite{Dodelson}. The
approximate analytical formula yields results for the small
angular scale acoustical oscillations that agree with reasonable
accuracy with the WMAP, Archeops  and Boomerang
data~\cite{WMAP,Archeops,Boomerang}. Since the phion condensates
do not couple to radiation and only couple with weak gravitational
strength to baryons, their density perturbations are not subject
to dissipation like photon and neutrino density perturbations
before recombination.

For large-scale cosmology, the phion condensate density on the
average dominates the matter density, while for {\it locally bound
galactic systems and clusters of galaxies} visible baryon density
dominates the matter density, and the MOG field equations and the
quantum BEC equations lead to an effective modification of the
Newtonian acceleration law for weak fields that yields an
excellent fit to the galaxy rotation curve data and the X-ray
cluster mass profile data~\cite{Brownstein,Brownstein2}.

\section{Classical Field Equations and Test Particle Motion}

The relativistic gravitational field equations in MOG are given by
(in units with $c=1$)~\cite{Moffat2}:
\begin{equation}
\label{Einsteineqs}G_{\mu\nu}-g_{\mu\nu}\Lambda+Q_{\mu\nu}=8\pi
GT_{\mu\nu},
\end{equation}
\begin{equation}
\label{Bequation} \nabla_\nu B^{\mu\nu}+\frac{\partial
V(\phi)}{\partial\phi_\mu}+\frac{1}{\omega}\nabla_\nu\omega
B^{\mu\nu}=-\frac{1}{\omega}J^\mu,
\end{equation}
where $G_{\mu\nu}=R_{\mu\nu}-\frac{1}{2}g_{\mu\nu}R$, $\Lambda$ is
the cosmological constant and $\nabla_\mu$ denotes covariant
differentiation with respect to the metric $g_{\mu\nu}$. We have
\begin{equation}
\label{Qequation} Q_{\mu\nu}=G(\nabla^\alpha\nabla_\alpha\Theta
g_{\mu\nu}-\nabla_\mu\nabla_\nu\Theta),
\end{equation}
where $G(x)$ is the scalar field spacetime dependent gravitational
``constant'' and $\Theta(x)=1/G(x)$. Moreover,
\begin{equation}
B_{\mu\nu}=\partial_\mu\phi_\nu-\partial_\nu\phi_\mu,
\end{equation}
and $J^\mu$ denotes a matter current. The potential $V(\phi)$ has
the form
\begin{equation}
V(\phi)=-\frac{1}{2}\mu^2\phi^\mu\phi_\mu+W(\phi),
\end{equation}
where $\mu(x)$ is the effective spacetime dependent mass of the
vector field $\phi^\mu$ and $W(\phi)$ is a $\phi^\mu$
self-interaction potential. The coupling constant $\omega(x)$ is
the scalar spacetime dependent coupling of the skew field
$B^{\mu\nu}$ to matter.

The total energy-momentum tensor is given by
\begin{equation}
T_{\mu\nu}=T_{M\mu\nu}+T_{\phi\mu\nu}+T_{S\mu\nu},
\end{equation}
where $T_{M\mu\nu}, T_{\phi\mu\nu}$ and $T_{S\mu\nu}$ denote the
energy-momentum contributions from the matter fields, the vector
field $\phi^\mu(x)$, and the scalar fields $G(x),\omega(x)$ and
$\mu(x)$, respectively.

From the Bianchi identities
\begin{equation}
\nabla_\nu G^{\mu\nu}=0,
\end{equation}
and from the field equations (\ref{Einsteineqs}), we obtain the
conservation law:
\begin{equation}
\label{conservation} \nabla_\nu T^{\mu\nu}+\frac{1}{G}\nabla_\nu
GT^{\mu\nu}-\frac{1}{8\pi G}\nabla_\nu Q^{\mu\nu}=0.
\end{equation}

The effective gravitational ``constant'' $G(x)$ satisfies the
field equations:
\begin{equation}
\label{Gequation} \nabla_\alpha\nabla^\alpha G+\frac{\partial
V(G)}{\partial G}+N=\frac{1}{2}G^2\biggl(T+\frac{\Lambda}{4\pi
G}\biggr),
\end{equation}
where
\begin{equation}
N=-3\Theta\biggl(\frac{1}{2}\nabla_\alpha G\nabla^\alpha G
+V(G)\biggr)+G\biggl(\frac{1}{2}\nabla_\alpha\omega\nabla^\alpha\omega
-V(\omega)\biggr)
$$ $$
+\frac{G}{\mu^2}\biggl(\frac{1}{2}\nabla_\alpha
\mu\nabla^\alpha\mu-V(\mu)\biggr)+\frac{3G^2}{16\pi}\nabla_\alpha\nabla^\alpha\Theta,
\end{equation}
and $T=g^{\mu\nu}T_{\mu\nu}$. The scalar field $\omega(x)$ obeys
the field equations
\begin{equation}
\label{omegaequation}
\nabla_\alpha\nabla^\alpha\omega+\frac{\partial
V(\omega)}{\partial\omega}+F=0,
\end{equation}
where
\begin{equation}
F=-\Theta\nabla_\alpha G\nabla^\alpha\omega
+G\biggl(\frac{1}{4}B^{\mu\nu}B_{\mu\nu}+V(\phi)\biggr).
\end{equation}
The field $\mu(x)$ satisfies the equations
\begin{equation}
\label{muequation} \nabla_\alpha\nabla^\alpha\mu+\frac{\partial
V(\mu)}{\partial\mu}+P=0,
\end{equation}
where
\begin{equation}
P=-\biggl[\Theta\nabla^\alpha G
\nabla_\alpha\mu+\frac{2}{\mu}\nabla^\alpha\mu\nabla_\alpha\mu
+\omega\mu^2 G\frac{\partial V(\phi)}{\partial\mu}\biggr],
\end{equation}
and the last term arises from the $\mu$ dependence of $V(\phi)$.

The equation of motion for a test particle of mass $m$ and charge
$\lambda$ when $\omega=\omega_{\rm ren}={\rm constant}$, where
$\omega_{\rm ren}$ is the renormalized value of $\omega$, is given
by
\begin{equation}
\frac{d^2x^\mu}{d\tau^2}+\Gamma^\mu_{\alpha\beta}\frac{dx^\alpha}{d\tau}\frac{dx^\beta}{d\tau}
=f^\mu,
\end{equation}
where $\tau$ is the proper time along the test particle trajectory
and
\begin{equation}
f^\mu=\frac{\lambda\omega}{m}{B^\mu}_\nu\frac{dx^\nu}{d\tau}.
\end{equation}
For massless photons the test charge $\lambda$ is zero and photons
move along null geodesics:
\begin{equation}
\frac{d^2x^\mu}{d\tau^2}
+\Gamma^\mu_{\alpha\beta}\frac{dx^\alpha}{d\tau}\frac{dx^\beta}{d\tau}=0.
\end{equation}

\section{Bose-Einstein Condensate of Phions}

An important prediction of quantum statistical mechanics is that
of a phase transition in an ideal gas of identical bosons when the
thermal de Broglie wavelength exceeds the mean spacing between
bosons~\cite{Parkins}. The bosons in the lowest energy state are
stimulated by the presence of other bosons to occupy the same
state, resulting in a macroscopic occupation of a single quantum
state that constitutes a macroscopic quantum-mechanical
system~\cite{Bose,Einstein}.

Well before recombination, the boson-phion particle can collapse
into the lowest energy quantum ground state when the temperature
of the phion fluid is below a critical temperature, $T < T_c$, and
form a degenerate Bose-Einstein fluid that does not couple to
photons and is not dissipated by photon
pressure~\cite{Ruffini,Sin,Hu2,Silverman,Ferrer}. The fluid has
zero viscosity and pressure and forms the dominant energy density
of matter in the universe at large cosmological scales greater
than the scales of galaxies and clusters of galaxies. The phion
Bose-Einstein condensates (BEC) are the dark matter contribution
to the cosmological density, which dominates the baryon density,
$\Omega_\phi > \Omega_b$ where $\Omega_\phi$ denotes the
fractional BEC density, at cosmological scales and couple weakly
with gravitational strength to baryon matter.

The phions undergo a second-order phase transition through a
spontaneous symmetry breaking mechanism below a critical
temperature $T < T_c$. This mechanism can generate the effective
phion mass $m_\phi$ and the BEC exhibits a broken symmetry and a
phase coherence. The condensate wave function is non-vanishing and
the ground state (vacuum) depends coherently on the phases of the
phions, such that they are spatially correlated throughout the
system. For a {\it quantum state} of light phions, it is possible
to form a non-relativistic quantum BEC of phions in a second order
phase transition.

The gravitational collapse (clumping) of the phion condensate
fluid to form gravitational potential wells before recombination,
allows the early formation of structures at small horizon scales
of the order of proto-galaxy sizes. These structures eventually
lead to the formation of clusters of galaxies in a hierarchical
scheme of structure growth. For this to happen, the density of
non-relativistic BEC phions $\rho_\phi$ dominates the clumping
fluid with a sub-horizon size of order of the size of a
proto-galaxy and with a suitable value of the coherence length
$R_c$.

A Jean's scale $\lambda_J$ separates gravitationally stable and
gravitationally non-stable modes. The time scale for gravitational
collapse is given by $\tau_{\rm grav}\sim 1/\sqrt{G_{\rm
ren}\rho_\phi}$. On the other hand, the time scale for fluid
pressure to dynamically respond is $\tau_{\rm press}\sim
\lambda/c_s$, where $\lambda$ is the size of the condensate in the
fluid and $c_s$ is the sound velocity. In units with $\hbar$ and
$c$ not equal to unity, for quantum non-relativistic BEC matter,
the velocity $v_c$ in the condensate fluid is determined by the de
Broglie wavelength $\lambda_B=h/m_\phi v_c$. The fraction of
phions in the coherent BEC ground state is about $100\%$ before
decoupling at a redshift $z_d\sim 1100$, and after decoupling
baryonic structures can begin to form and the quantum Jeans scale
$\lambda_{\rm Jquant}$ can become the size of proto-galaxies.

Three characteristic lengths define the system: the Compton
wavelength, $\lambda_c=h/m_\phi c$, the average distance between
particles, $\lambda_d=n_\phi^{-1/3}$ where $n_\phi$ denotes the
number density of phions, and the thermal wavelength
$\lambda_T=h/p$. The thermal de Broglie wavelength is
\begin{equation}
\lambda_T=\biggl(\frac{2\pi \hbar^2}{m_\phi k_BT}\biggr)^{1/2},
\end{equation}
where $k_B$ is Boltzmann's constant. The non-relativistic limit
corresponds to $\lambda_T\gg \lambda_c$, while the relativistic
limit holds for $\lambda_T\ll \lambda_c$.

For an ideal non-relativistic Bose gas, the number of phions is
given by ~\cite{Huang,Pathria}:
\begin{equation}
N_\phi\equiv \sum_{\epsilon_\phi}\langle
n_{\epsilon_\phi}\rangle=\sum_{\epsilon_\phi}\frac{1}{z^{-1}\exp(\epsilon_\phi/k_BT)-1},
\end{equation}
where $\epsilon_\phi$ denotes the phion energy spectrum and $z$ is
the fugacity, related to the chemical potential $\mu_{\rm chem}$
by $z=\exp(\mu_{\rm chem}/k_BT)$. The fugacity $z$ is solved in
terms of the equation
\begin{equation}
n_\phi=\frac{2\pi}{h^3}(2m_\phi)^{3/2}\int_0^\infty
d\epsilon_\phi\frac{\epsilon^{1/2}_\phi}
{z^{-1}\exp(\epsilon_\phi/k_BT)-1}+\frac{1}{V}\frac{z}{1-z},
\end{equation}
where $V$ is a characteristic volume. We can solve this equation
to obtain
\begin{equation}
\label{nequation}
n_\phi=\frac{g_{3/2}(z)}{\lambda^3_T}+\frac{1}{V}\frac{z}{1-z},
\end{equation}
where
\begin{equation}
g_n(z)=\frac{1}{\Gamma(n)}\int dx\frac{x^{n-1}}{z^{-1}\exp(x)-1}.
\end{equation}
We can rewrite (\ref{nequation}) as
\begin{equation}
\lambda^3_T\frac{\langle n_{0\phi}\rangle}{V}=\lambda^3_T
n_\phi-g_{3/2}(z),
\end{equation}
which implies that $\langle n_{0\phi}\rangle/V > 0$ when the
temperature $T$ and $n_\phi$ are such that
\begin{equation}
\lambda^3_\phi n_\phi > \zeta(3/2)=2.612...,
\end{equation}
where $g_{3/2}(1)=\zeta(3/2)$ and $\zeta(x)$ is the Riemann-zeta
function. A finite number of phion bosons occupy the ground state
level with $\epsilon_\phi=0$. This defines the critical
temperature for phion condensation:
\begin{equation}
\label{criticalT} T_c=\frac{2\pi\hbar^2
n_\phi^{2/3}}{k_Bm_\phi\zeta^{2/3}(3/2)},
\end{equation}
and the critical phion number density
\begin{equation}
n_{c\phi}=\frac{\zeta(3/2)}{\lambda^3_T}.
\end{equation}
Thus, for $T < T_c$ a large number of phions occupy the ground
state $\epsilon_\phi=0$.

We are required in a relativistic cosmology to take into account
the expansion of the universe. The solution for the case of an
ideal boson gas in a Friedmann-Robertson-Walker (FRW) universe is
complicated, so
authors~\cite{Schrodinger,Altaie,Parker,Ford,Aragao} have
considered a boson gas in a closed Einstein universe. In  the
spatial geometry of a 3-sphere, the metric takes the form:
\begin{equation}
ds^2=c^2dt^2-a_E^2[d\chi^2+\sin^2\chi(d\theta^2+\sin^2\theta
d\phi^2)],
\end{equation}
where $\chi$ and $\theta$ run from $0$ to $2\pi$. The radius of
the Einstein universe $a_E$ is constant. The solution for the spin
1 phion discrete energy spectrum is given by ($\hbar=c=1$):
\begin{equation}
\epsilon_n=\frac{1}{a_E}(n^2+m^2_\phi a^2_E)^{1/2}
\end{equation}
with degeneracy $d_n=2(n^2-1)$ where $n=2,3,...$.

Let us take the mean phion density at large scale galactic
distances to be of the order of the critical density,
$\rho_\phi\sim 9.52\times 10^{-30}\,g\,{\rm cm}^{-3}$. From
(\ref{criticalT}), we get in the condensation region
\begin{equation}
n_\phi \leq
\biggl(\frac{k_BT_c\rho_\phi}{2\pi\hbar^2\zeta^{2/3}(3/2)}\biggr)^{3/5}.
\end{equation}
We obtain for the phion condensate density that is comparable to
the cosmological critical density for a critical temperature at
recombination $T\sim 10^5\,K$:
\begin{equation}
n_\phi \leq 6.3\times 10^7\,{\rm cm}^{-3}.
\end{equation}
We get from $m_\phi=\rho_\phi/n_\phi$ for the {\it effective}
phion condensate mass at the epoch of recombination:
\begin{equation}
m_\phi\sim 8.5\times 10^{-5}\,eV/c^2.
\end{equation}

\section{Bose-Einstein Gas and the Ground State}

A Bose-Einstein state was experimentally verified by the Joint
Institute for Laboratory Astrophysics (JILA) in 1995. Anderson et
al.~\cite{Anderson} produced a condensate of spin-polarized
$^{87}RB$ atoms in a confining magnetic trap. Densities and
temperatures of order $10^{11}\,{\rm cm^{-3}}$ and tens of
micro-Kelvin, respectively, were achieved.

We shall assume that the phion boson inter-particle interactions
are weak, so that at low temperatures in cosmology the de Broglie
wavelengths of the phions are large compared to the inter-phion
separation. The phion-phion interactions are dominated by elastic
S-wave scattering and we can consider for weak fields of
gravitational strength only two-body collisions. The scattering
length can be positive, $a > 0$, for a repulsive interaction and
negative, $a < 0$, for an attractive interaction. The phion
self-interaction potential is given by
\begin{equation}
U({\bf x}-{\bf x}')=g\delta({\bf x}-{\bf x}').
\end{equation}
Here, the strength of the interaction is
\begin{equation}
g=\frac{4\pi\hbar^2a}{m_\phi}.
\end{equation}
The Hamiltonian for weakly interacting phions in an external
gravitational potential $V_{\rm grav}$ is of the form
\begin{equation}
H=\int d^3x\psi^\dagger({\bf
x})\biggl[-\frac{\hbar^2}{2m_\phi}{\vec \nabla}^2+V_{\rm
grav}({\bf x})\biggr]\psi({\bf x})
$$ $$
+\frac{1}{2}\int d^3x\int d^3 x'\psi^\dagger({\bf
x})\psi^\dagger({\bf x}')U({\bf x}-{\bf x}')\psi({\bf
x}')\psi({\bf x}),
\end{equation}
where $\psi({\bf x})$ and $\psi^\dagger({\bf x})$ are the phion
field annihilation and creation operators, respectively.

A necessary condition for the applicability of a calculation of
the BEC gas is that $a^3\rho_\phi \ll 1$. The equation of motion
for the wavefunction takes the form
\begin{equation}
\label{Heisenbergeq} i\hbar\frac{\partial\psi({\bf x},t)}{\partial
t}=\biggl[-\frac{\hbar^2}{2m_\phi}{\vec\nabla}^2+V_{\rm grav}({\bf
x})\biggr]\psi({\bf x},t)+g\psi^\dagger({\bf x},t)\psi({\bf
x},t)\psi({\bf x},t).
\end{equation}
To solve this equation, a mean-field approach is adopted in which
\begin{equation}
\psi({\bf x},t)={\overline\psi}({\bf x},t)+\delta\psi({\bf x},t),
\end{equation}
where ${\overline\psi}=\langle\psi\rangle$ is the phion condensate
wave function and $\delta\psi$ is the thermal quantum fluctuation
around the mean value. The quantity $\overline\psi$ is an ``order
parameter'' that describes a broken gauge symmetry; the Abelian
gauge symmetry occurs in the classical MOG action for a {\it
massless} phion field, $m_\phi=0$~\cite{Moffat2}. Moreover, we
have that $\langle\delta\psi\rangle=0$ and for small $\delta\psi$
the temperature is assumed to be $T\sim 0$.

The basic equation for the condensate phion system is the
Gross-Pitaevskii equation~\cite{Gross,Pitaevskii,Lifshitz}:
\begin{equation}
\label{Gross} i\hbar\frac{\partial\psi({\bf x},t)}{\partial
t}=\biggl[-\frac{\hbar^2}{2m_{\phi}}{\vec\nabla}^2+V_{\rm
grav}({\bf x})+N_{\phi} g\vert\psi({\bf
x},t)\vert^2\biggr]\psi({\bf x},t).
\end{equation}
A stationary solution for the phion wave function can be obtained
from
\begin{equation}
\psi({\bf x},t)=\exp(-i\mu_{\rm chem}t/\hbar)\tilde\psi({\bf x}).
\end{equation}
This leads to the time-independent equation
\begin{equation}
\label{GPtind}
\biggl[-\frac{\hbar^2}{2m_\phi}{\vec\nabla}^2+V_{\rm grav}({\bf
x})+N_{\phi} g\vert{\tilde\psi({\bf
x})}\vert^2\biggr]\tilde\psi({\bf x})=\mu_{\rm
chem}\tilde\psi({\bf x}).
\end{equation}

The phion condensate gas before recombination can become unstable
for a negative scattering length, $a < 0$, corresponding to an
attractive inter-phion potential. A stable local minimum for the
energy exists only up to a maximum number of phion condensates,
and above this number the kinetic energy can no longer stabilize
the condensate gas against collapse. This instability describes
the early formation of proto-galaxies before recombination and
growth of structure after recombination. In practise, the energy
per phion condensate has three contributions: the kinetic energy,
the phion potential energy and the potential energy due to
gravity. An application of the virial theorem to the proto-galaxy
condensate gas gives
\begin{equation}
{\cal E}_{\rm kin}={\cal E}_{\phi}-\frac{3}{2}{\cal E}_{\rm grav},
\end{equation}
where ${\cal E}_{\rm kin}, {\cal E}_{\phi}$ and ${\cal E}_{\rm
grav}$ denote the kinetic energy, the phion potential energy and
the gravitational potential energy, respectively. As the number
$N_\phi$ of phion condensates increases, a critical point is
reaches when the condensate gas collapses and clumping can begin
and form a proto-galaxy. The repulsive phion-phion forces can take
over at a certain stage of the gravitational collapse of the phion
condensate gas for a scattering length $a > 0$, preventing the
formation of numerous unobserved compact sub-structures in N-body
simulations.

The critical chemical potential for a transition is given by
\begin{equation}
\mu_{c{\rm chem}}=\frac{8\pi a\hbar^2}{m_\phi\lambda^3}\zeta(3/2),
\end{equation}
where $a$ is the scattering length. For a galaxy-sized object, we
can choose $a\sim\lambda_G\sim 14\,{\rm kpc}$ and $m_\phi\sim
10^{-24}\,eV/c^2$ to give
\begin{equation}
\frac{\mu_{c{\rm chem}}}{\hbar}\sim 10^{-14}\, s^{-1},
\end{equation}
corresponding to a period $t_p\sim 10^{14}\, s$ which is roughly
the time it takes for a light signal to cross a galaxy, $R_G/c\sim
10^{12}\, s$.

\section{Phion BEC and Evolution of the Universe}

In the late-time universe, the non-relativistic phion matter {\it
locally} in bound systems such as stars, galaxies and clusters of
galaxies is dominated by baryon matter and neutral hydrogen and
helium gases. The quantum non-relativistic, phion condensate is
subjected to only the weak fifth force with baryon matter, so that
as the universe expands there is little or no decoherence of the
phion condensate gas. However, during the evolution of galaxies
ordinary baryonic matter begins to dominate inside the cores of
galaxies and traces light. This predicts that {\it cold dark
matter halos do not exist around the cores of galaxies or in
clusters of galaxies}. On the other hand, at inter-galactic
cosmological scales the neutral BEC phions dominate the density of
matter with $\Omega_b < \Omega_\phi$. The spontaneous symmetry
breaking leading to the phion BEC is relaxed inside the cores of
galaxies, so that the effective mass $m_\phi$ of the phions is
significantly reduced, $m_\phi\sim 10^{-24}$ eV, corresponding to
a Compton wavelength $\lambda_c=1/\mu_\phi=h/m_\phi c\sim 14$ kpc.

Locally, inside the bound galaxy systems well after decoupling and
the formation of galaxies, the spherically symmetric, static
solution of the classical MOG field equations, together with the
variation and renormalization of Newton's gravitational constant
$G$, leads at late times to the {\it effective} modification of
Newtonian acceleration for weak fields in galaxies~\cite{Moffat2}.
The expression for the modified acceleration is obtained from the
static spherically symmetric solution of the equation valid
outside the matter source
\begin{equation}
\label{phioneq} \phi''_0+\frac{2}{r}\phi'_0-\mu_\phi^2\phi^0=0,
\end{equation}
where $'$ denotes differentiation with respect to the radial
coordinate $r$ and $\phi_0$ denotes the time component of the
phion field $\phi_\mu$. The solution to (\ref{phioneq}) is the
Yukawa potential and the modified acceleration law takes the form:
\begin{equation}
\label{modifiedNewton} a(r)=-\frac{G_\infty
M}{r^2}+K\frac{\exp(-\mu_\phi r)}{r^2}(1+\mu_\phi r),
\end{equation}
where
\begin{equation}
K=G_N\sqrt{M}\sqrt{M_0}.
\end{equation}
Here, $G_N$ denotes Newton's gravitational constant, $M$ is the
total mass of the galaxy and $M_0$ is a constant that measures the
strength of the coupling of the skew field $B^{\mu\nu}$ to matter,
expressed in units of the square-root of a mass. We choose
\begin{equation}
G_\infty=G_N(1+\alpha),
\end{equation}
where
\begin{equation}
\alpha=\sqrt\frac{M_0}{M}.
\end{equation}

We obtain the expression for the {\it effective} modified
Newtonian acceleration law:
\begin{equation}
\label{modaccel} a(r)=-\frac{G_{\rm
ren}M}{r^2}[1+\alpha(1-\exp(-\mu_\phi r)(1+\mu_\phi r))].
\end{equation}
We can express the acceleration in terms of an effective Newtonian
law in the form:
\begin{equation}
a(r)=-\frac{G_{\rm eff}(r)M(r)}{r^2},
\end{equation}
where
\begin{equation}
G_{\rm eff}(r)= G_N[1+\alpha(1-\exp(-\mu_\phi r)(1+\mu_\phi r))].
\end{equation}

We note that the integration constant $K$, $\alpha$ and $G_{\rm
eff}$ depend on the inverse square root of the total mass $M$.
This non-linear dependence on the mass $M$ cannot be interpreted
in terms of a classical collection of point masses. However, our
classical modification of Newton's acceleration law is a
consequence of an effective classical limit of the non-linear
Schr\"odinger equation for the phion BEC. Let us consider an
application of Ehrenfest's theorem:
\begin{equation}
\frac{d}{dt}\langle {\bf x}\rangle=\langle {\bf p}/m\rangle,\quad
\frac{d}{dt}\langle {\bf p}/m\rangle=\langle {\bf a}\rangle.
\end{equation}
The Hamiltonian operator obtained from Eq.(\ref{Gross}) is given
by
\begin{equation}
{\hat H}=-\frac{\hbar^2}{2m_\phi}{\vec\nabla}^2+V_{\rm grav}({\bf
x})+W({\bf x}),
\end{equation}
where for a static potential
\begin{equation}
W({\bf x})=N_{\phi} g\vert\tilde\psi({\bf x})\vert^2,
\end{equation}
and $-i\hbar{\vec\nabla}={\bf p}$. We get
\begin{equation}
\frac{d}{dt}\langle{\bf p}\rangle=\frac{i}{\hbar}\langle [H,{\bf
p}]\rangle =-(\langle{\vec\nabla}V_{\rm grav}({\bf x})
\rangle+\langle{\vec\nabla}W({\bf x})\rangle),
\end{equation}
where $-\langle{\vec\nabla}V_{\rm grav}\rangle$ is the Newtonian
force law and $-\langle{\vec\nabla}W\rangle$ is the quantum, BEC
phion modification of the Newtonian force law corresponding to the
modified acceleration law in (\ref{modifiedNewton}).

The effective modified acceleration law (\ref{modaccel}) fits
remarkably well the rotation curve data of galaxies and the mass
profiles of X-ray clusters~\cite{Brownstein,Brownstein2}. This
avoids the problem of standard cold dark matter models that
predict density profiles within galactic cores that are too sharp
and ``cuspy''~\cite{deBlok}. The LSB galaxies have a small
contribution from the baryonic mass component, so they are
efficient tracers of rotation curves, and can help to distinguish
dark matter halo models from the modified Newtonian acceleration
law in MOG. The velocity rotation curves predicted by MOG fit well
the inner cores of dwarf galaxies and LSBs in term of one
parameter, the mass-to-light ratio $<M/L>$, using photometric
data~\cite{Brownstein}.

The MOG acceleration formula predicts that for satellite galaxies
the rotational velocities and density profiles of the satellites
have a Newtonian-Kepler behavior.

\section{Weak Field Equations in Cosmology}

In a homogeneous universe with small perturbations, the metric is
given in a spatially flat universe in the conformal
Newtonian-Poisson gauge by
\begin{equation}
ds^2=a^2[(1+2\Phi)d\eta^2-(1-2\Phi)\delta_{ik}dx^idx^k],
\end{equation}
where $\eta$ denotes the conformal time and $\Phi\ll 1$.

For weak gravitational fields the Poisson equation for the
gravitational potential $\Phi$ takes the form:
\begin{equation}
{\vec\nabla}^2\Phi({\bf x},t)+a^2(t)Y({\bf x},t)=4\pi a^2(t)G({\bf
x},t)\delta\rho({\bf x},t),
\end{equation}
where $Y$ denotes the contributions obtained from
(\ref{Qequation}), the factor $a^2(t)$ is inserted to account for
the difference between the FRW co-moving coordinate vector ${\bf
x}$ and the coordinate vector $a(t){\bf x}$ that measures the
proper distances at time $t$, and $\delta\rho$ denotes the size of
the density fluctuation. If we assume that for large cosmological
scales, we can ignore the $Y$ contributions and that $G=G_{\rm
ren}={\rm constant}$, then we obtain~\cite{Weinberg}:
\begin{equation}
\label{Phiequation} \Phi({\bf x},t)=-4\pi G_{\rm
ren}t^{2/3}a^2(t)\int d^3q\frac{1}{q^2}\exp({i{\bf q}\cdot{\bf
x}})N_{\bf q},
\end{equation}
where $N_{\bf q}=\delta_{\bf q}/t^{2/3}$, $\delta_{\bf
q}=\delta\rho_{\bf q}/\rho_{\bf q}$. The co-moving wave number
vector is denoted by ${\bf q}$, which is related to the physical
wave number at last scattering ${\bf k}={\bf q}/a(t_{LS})$ where
$t_{LS}$ denotes the time of last scattering.

The equation for the vector field $\phi_\mu$ for weak fields is
given by~\cite{Moffat2}:
\begin{equation}
{\vec\nabla}^2\phi_\mu({\bf x},t)-a^2(t)\mu_\phi^2\phi_\mu({\bf
x},t)-a^2(t)\frac{\partial W(\phi)({\bf
x},t)}{\partial\phi^\mu({\bf
x},t)}=a^2(t)\frac{1}{\omega}J_\mu({\bf x},t),
\end{equation}
where we have assumed that the coupling constant $\omega$ and the
mass $\mu_\phi$ have their renormalized values, and
$J^\mu=(q,J^i)\,(i=1,2,3)$ is a matter current.

\section{Angular Distance and Cosmological Parameters}

Before recombination the photons couple strongly to baryons, and
after recombination hydrogen gas becomes neutral and the photons
no longer interact with baryons. The free photons move along null
geodesics and are described by the distribution function $f$:
\begin{equation}
dN=f(x^i,p_j,\eta)d^3xd^3p,
\end{equation}
where $dN$ denotes the number of photons at time $\eta$. In the
absence of scattering, the distribution function $f$ obeys the
collisionless Boltzmann equation
\begin{equation}
\frac{\partial f}{\partial\eta}+\frac{dx^i}{d\eta}\frac{\partial
f}{\partial x^i}+\frac{dp^i}{d\eta}\frac{\partial f}{\partial
p^i}=0.
\end{equation}

The geodesic equations for the photons are
\begin{equation}
\frac{dx^\mu}{d\tau}=p^\mu,\quad
\frac{dp_\mu}{d\tau}=\frac{1}{2}\frac{\partial
g_{\gamma\delta}}{\partial x^\mu}p^\gamma p^\delta.
\end{equation}
The Boltzmann equation takes the form
\begin{equation}
\frac{\partial f}{\partial\eta}+n^i(1+2\Phi)\frac{\partial
f}{\partial x^i}+2p\frac{\partial\Phi}{\partial x^j}\frac{\partial
f}{\partial p_j}=0,
\end{equation}
where $n_i=-p_i/p$ is a vector that determines the direction on
the sky from which the photons reach an observer.

Under general assumptions, the fractional variation from the mean
of the CMB temperature observed in the direction ${\hat {\bf n}}$
takes the form~\cite{Weinberg}:
\begin{equation}
\frac{\Delta T}{T}({\hat{\bf n}})=\int d^3k\delta\rho_{\bf
k}\exp({id_A{\hat{\bf n}}\cdot{\bf k}})\biggl[{\cal F}(k)+i{\hat
{\bf n}}\cdot{\hat {\bf k}}{\cal G}(k)\biggr],
\end{equation}
where $d_A$ is the angular diameter distance from the surface of
last scattering, and $k^2\delta\rho_{\bf k}$ is proportional to
the Fourier transform of the fractional total energy perturbation.
The form factor ${\cal F}(k)$ arises from the Sachs-Wolfe effect
and the intrinsic temperature fluctuations, while ${\cal G}(k)$
arises from the Doppler effect.

The angular diameter distance of the surface of last scattering is
given by
\begin{equation}
d_A=\frac{1}{\Omega_c^{1/2}H_0(1+z_L)}\sinh\biggl[\Omega_c^{1/2}\int^1_{1/(1+z_L)}
\frac{dx}{(\Omega_\Lambda
x^4+\Omega_kx^2+\Omega_mx)^{1/2}}\biggr],
\end{equation}
where $z_L\sim 1100$ is the redshift at the surface of last
scattering, $\Omega_k=1-\Omega_\Lambda-\Omega_m$ is the curvature
parameter, and $\Omega_m$ and $\Omega_\Lambda$ denote
\begin{equation}
\Omega_m=\frac{8\pi G_{\rm ren}\rho_m}{3H^2},\quad
\Omega_\Lambda=\frac{\Lambda}{3H^2}.
\end{equation}
Here, $G_{\rm ren}$ is the renormalized value of the gravitational
constant determined at some time ${\bar t}$ before the time of
recombination $t\sim t_r$:
\begin{equation}
\label{Grenorm} G_{\rm ren}=G_0(1+Z),
\end{equation}
where $G_0\sim G_N$ is the ``bare'' Newtonian value of the
gravitational constant. In our modified gravitational theory, the
density of matter is given by
\begin{equation}
\Omega_m=\Omega_b+\Omega_\phi+\Omega_S,
\end{equation}
where $\Omega_\phi$ and $\Omega_S$ are the densities of matter
associated with the phion BEC density and the scalar fields,
$G,\omega$ and $\mu$, respectively.

\section{Baryon-Radiation Fluid and Phion BEC Fluid Before Recombination}

The fluctuations of the background radiation depend upon the
radiation energy density fluctuations,
$\delta\rho_\gamma/\rho_\gamma$, and the gravitational potential
$\Phi$. The medium consists of two components, the coupled
baryon-radiation fluid and the non-relativistic phion BEC
condensate fluid and scalar fields fluid. The light phion
condensates are {\it electrically neutral and do not couple to
photons} and couple to baryons with a gravitational strength. We
shall assume that the phion BEC condensate density fluctuations
$\delta_{\phi}=\delta\rho_{\phi}/\rho_{\phi}$ dominate i.e.,
$\delta_{\phi} \gg \delta_b$ and $\delta_{\phi} \gg \delta_S$,
where $\delta_b=\delta\rho_b/\rho_b$ and
$\delta_S=\delta\rho_S/\rho_S$ denote the baryon and scalar field
fluctuations of the neutral scalar fields $G, \omega$ and $\mu$,
respectively.

For an imperfect fluid with energy density $\rho$ and pressure
$p$, the conservation law for the standard energy-momentum tensor
is $\nabla_\alpha{T^\alpha}_\beta=0$, which leads to the first
order perturbation equation for the components
$\nabla_\alpha{T^\alpha}_0=0$~\cite{Mukhanov}:
\begin{equation}
\label{0conservation} \delta\rho'+3H(\delta\rho+\delta
p)-3(\rho+p)\Phi'+a(\rho+p)\partial_iu^i=0,
\end{equation}
where $'$denotes differentiation with respect to $\eta$, $u^i$
denotes the spatial velocity for $i=1,2,3$. Another equation is
obtained from the components of conservation,
$\nabla_\alpha{T^\alpha}_i=0$:
\begin{equation}
\frac{1}{a^4}[a^5(\rho+p)\partial_iu^i]'-\frac{4}{3}\eta_{\rm
vis}\Delta\partial_iu^i+\Delta\delta p+(\rho+p)\Delta\Phi=0,
\end{equation}
where $\eta_{\rm vis}$ is the shear viscosity. These two equations
hold {\it separately} for the baryon-radiation fluid and the
neutral phion BEC fluid.

For the phion condensate density $\rho_\phi$, {\it the classical
pressure $p$ and the shear viscosity $\eta_{\rm vis}$ are equal to
zero}. From (\ref{0conservation}) and $\rho_{\phi}a^3={\rm
const.}$, we obtain
\begin{equation}
(\delta_{\phi}-3\Phi)'+a\partial_iu^i=0.
\end{equation}
The dissipation caused by photon pressure and shear viscosity
$\eta_{\rm vis}$ cannot be neglected for the baryon-radiation
fluid and leads to Silk damping erasure of the baryon-radiation
perturbations at recombination.

For non-relativistic baryons, the energy conservation law,
$\nabla_\alpha{T^\alpha}_0=0$, reduces to a conservation of total
baryon number and we obtain from (\ref{0conservation}):
\begin{equation}
(\delta_b-3\Phi)'+a\partial_iu^i=0.
\end{equation}
The equation for the perturbations in the radiation component,
$\delta_\gamma=\delta\rho_\gamma/\rho_\gamma$, is given by
\begin{equation}
(\delta_\gamma-4\Phi)'+\frac{4}{3}a\partial_iu^i=0.
\end{equation}
The tightly coupled baryon-photon fluid moves with a single
velocity, and we get
\begin{equation}
\frac{\delta s}{s}=\frac{3}{4}\delta_\gamma-\delta_b={\rm const.},
\end{equation}
where $\delta s/s$ denotes the fractional entropy fluctuations in
the fluid. Adiabatic perturbations give $\delta s=0$ and
$\delta_b=(3/4)\delta_\gamma$.

The speed of sound in the baryon-photon fluid is given by
\begin{equation}
c^2_s\equiv \frac{\delta p}{\delta\rho}=\frac{\delta
p_\gamma}{\delta\rho_\gamma+\delta\rho_b}
=\frac{1}{3}\biggl(1+\frac{3}{4}\frac{\rho_b}{\rho_\gamma}\biggr)^{-1}.
\end{equation}
The shear viscosity $\eta_{\rm vis}$ is given by
\begin{equation}
\eta_{\rm vis}=\frac{4}{15}\rho_\gamma\tau_\gamma,
\end{equation}
where $\tau_\gamma$ is the photon mean-free time.

The dissipation scale for co-moving wave-number is
\begin{equation}
k_D(\eta)=\biggl(\frac{2}{5}\int_0^\eta d\eta
c^2_s\frac{\tau_\gamma}{a}\biggr)^{-1/2}.
\end{equation}
We see from this equation that the viscosity damps the
perturbations for co-moving scales $\lambda \leq 1/k_D$. For
$c_s^2=1/3$ and assuming instantaneous recombination, the
dissipation scale is given by~\cite{Mukhanov}:
\begin{equation}
\frac{1}{k_D\eta_r}\sim
0.6(\Omega_\phi)^{1/4}(\Omega_b)^{-1/2}(z_r)^{-3/4},
\end{equation}
where $\eta_r$ and $z_r$ denote the time and redshift at
recombination, respectively.

\section{Correlation Function and Multipoles}

The correlation function for the temperature differences over the
sky for a given angle $\theta$ is
\begin{equation}
C(\theta)=\left<\frac{\delta T}{T_0}({\bf n}_1)\frac{\delta
T}{T_0}({\bf n}_2)\right>,
\end{equation}
where $<..>$ denotes the average over all ${\bf n}_1$ and ${\bf
n}_2$ with ${\bf n}_1\cdot{\bf n}_2=\cos\theta$ and the monopole
and dipole contributions have been subtracted. We write the
correlation function in the form
\begin{equation}
C(\theta)=\frac{1}{4\pi}\sum^\infty_{l=2}(2l+1)C_lP_l(\cos\theta),
\end{equation}
where
\begin{equation}
\label{CorrelationC} C_l=\frac{2}{\pi}\int
k^2dk\left|\biggl(\Phi(\eta_r)+\frac{\delta_k(\eta_r)}{4}\biggr)
j_l(k\eta_0)-\frac{3\delta_k'(\eta_r)}{4k}\frac{dj_l(k\eta_0)}{d(k\eta_0)}\right|^2,
\end{equation}
and $\eta_0$ denotes the present conformal time. The
$P_l(\cos\theta)$ and $j_l(k\eta)$ are the Legendre polynomials
and spherical Bessel functions, respectively. In terms of
spherical harmonics $C_l=\langle\vert a_{lm}\vert^2\rangle$ and
for $\theta < 1$ the dominant contribution is for $l\sim
1/\theta$.

In the standard calculations of the CMB power spectrum, the
results depend on the various cosmological parameters. The generic
inflation prediction is, $\vert\Phi^2_kk^3\vert=Bk^{n_s-1}$, with
$1-n_s\sim 0.03-0.08$. The amplitude $B$ is not predicted and has
to be fitted to the observations and we assume a flat spectrum
$n_s=1$. The total density of matter is given in our modified
gravity theory by
\begin{equation}
\Omega_m=\Omega_b+\Omega_\nu +\Omega_\phi+\Omega_S,
\end{equation}
where
\begin{equation}
\Omega_i=\frac{8\pi G_{\rm ren}\rho_i}{3H^2}.
\end{equation}
Here, $i$ denotes the contributions from the baryons, neutrinos,
phions and scalar field components. We have assumed that the
gravitational constant has reached a constant renormalized value
(\ref{Grenorm}) before recombination.

\section{Calculation of the Acoustic Oscillation Spectrum}

Mukhanov~\cite{Mukhanov} has obtained an analytical solution to
the amplitude of fluctuations for $l\gg 1$:
\begin{equation}
\label{fluctuations} l(l+1)C_l\sim \frac{B}{\pi}(O+N).
\end{equation}
Here, $O$ denotes the oscillating part of the spectrum, while the
non-oscillating contribution can be written as the sum of three
parts
\begin{equation}
N=N_1+N_2+N_3.
\end{equation}

The oscillating contributions can be calculated from the formula
\begin{equation}
O\sim
\sqrt{\frac{\pi}{r_hl}}\biggl[A_1\cos\biggl(lr_p+\frac{\pi}{4}\biggr)
+A_2\cos\biggl(2lr_p+\frac{\pi}{4}\biggr)\biggr]\exp(-(l/l_s)^2),
\end{equation}
where $r_h$ and $r_p$ are parameters that determine predominantly
the heights and positions of the peaks, respectively. The $A_1$
and $A_2$ are constant coefficients given in the range $100 < l <
1200$ for $\Omega_b\ll \Omega_\phi$ by
\begin{equation}
\label{A1coefficient} A_1\sim
0.1\xi\frac{((P-0.78)^2-4.3)}{(1+\xi)^{1/4}}\exp\biggl(\frac{1}{2}(l_s^{-2}-l_f^{-2})l^2\biggr),
\end{equation}
\begin{equation}
\label{A2coefficient} A_2\sim
0.14\frac{(0.5+0.36P)^2}{(1+\xi)^{1/2}},
\end{equation}
where
\begin{equation}
P=\ln\biggl(\frac{lI}{200(\Omega_\phi)^{1/2}}\biggr),
\end{equation}
and $I$ is given by the ratio
\begin{equation}
\frac{\eta_x}{\eta_0}\sim
\frac{I}{z_x^{1/2}}=3\biggl(\frac{\Omega_\Lambda}{\Omega_\phi}\biggr)^{1/6}
\biggl(\int_0^y\frac{dx}{(\sinh
x)^{2/3}}\biggr)^{-1}\frac{1}{z_x^{1/2}}.
\end{equation}
Here, $\eta_x$ and $z_x$ denote a time and a redshift in the range
$\eta_0 > \eta_x > \eta_r$ when radiation can be neglected and
$y=\sinh^{-1}(\Omega_\Lambda/\Omega_\phi)^{1/2}$. To determine
$\eta_x/\eta_0$, we use the exact solution for a flat
dust-dominated universe with a cosmological constant $\Lambda$:
\begin{equation}
a(t)=a_0\biggl(\sinh\biggl(\frac{3}{2}\biggr)H_0t\biggr)^{2/3},
\end{equation}
where $a_0$ and $H_0$ denote the present values of $a$ and the
Hubble parameter $H$. A numerical fitting formula
gives~\cite{Mukhanov}:
\begin{equation}
P\sim\ln\biggl(\frac{l}{200(\Omega_\phi^{0.09})(\Omega_\phi)^{1/2}}\biggr),\quad
r_p=\frac{1}{\eta_0}\int d\eta c_s(\eta).
\end{equation}
Moreover,
\begin{equation}
\xi\equiv
\frac{1}{3c_s^2}-1=\frac{3}{4}\biggl(\frac{\rho_b}{\rho_\gamma}\biggr),
\end{equation}
where
\begin{equation}
c_s(\eta)=\frac{1}{\sqrt{3}}\biggl[1+\xi\biggl(\frac{a(\eta)}{a(\eta_r)}\biggr)\biggr]^{-1/2}.
\end{equation}
For the matter-radiation universe:
\begin{equation}
a(\eta)={\bar
a}\biggl[\biggl(\frac{\eta}{\eta_{*}}\biggr)^2+2\biggl(\frac{\eta}{\eta_{*}}\biggr)\biggr],
\end{equation}
where for radiation-matter equality $z=z_{eq}$:
\begin{equation}
\frac{z_{eq}}{z_r}\sim
\biggl(\frac{\eta_r}{\eta_*}\biggr)^2+2\biggl(\frac{\eta_r}{\eta_*}\biggr),
\end{equation}
and $\eta_{eq}=\eta_*(\sqrt{2}-1)$ follows from ${\bar
a}=a(\eta_{eq})$.

The speed of sound at recombination depends only on the baryon
density, determining the deviation from its value in a purely
ultra-relativistic medium, and it can be expressed as
$c_s^2=1/3(1+\xi)$.

The $l_f$ and $l_s$ in (\ref{A1coefficient}) denote the finite
thickness and Silk damping scales, respectively, given by
\begin{equation}
l_f^2=\frac{1}{2\sigma^2}\biggl(\frac{\eta_0}{\eta_r}\biggr)^2,\quad
l_s^2=\frac{1}{2(\sigma^2+1/(k_D\eta_r)^2)}\biggl(\frac{\eta_0}{\eta_r}\biggr)^2,
\end{equation}
where
\begin{equation}
\sigma\sim 1.49\times 10^{-2}\biggl[1+\biggl(1+\frac{z_{\rm
eq}}{z_r}\biggr)^{-1/2}\biggr].
\end{equation}

For the non-oscillating parts, we have
\begin{equation}
N_1\sim
0.063\xi^2\frac{(P-0.22(l/l_f)^{0.3}-2.6)^2}{1+0.65(l/l_f)^{1.4}}\exp(-(l/l_f)^2),
\end{equation}
\begin{equation}
N_2\sim\frac{0.037}{(1+\xi)^{1/2}}\frac{P-0.22(l/l_s)^{0.3}+1.7)^2}{1+0.65(l/l_f)^{1.4}}\exp(-(l/l_s)^2),
\end{equation}
\begin{equation}
N_3\sim\frac{0.033}{(1+\xi)^{3/2}}\frac{P-0.5(l/l_s)^{0.55}+2.2)^2}{1+2(l/l_s)^2}\exp(-(l/l_s)^2).
\end{equation}

Mukhanov's formula for the oscillating spectrum is given by
\begin{equation}
\label{Mukhanov}  C(l)\equiv\frac{l(l+1)C_l}{[l(l+1)C_l]_{{\rm
low}\,l}}=\frac{100}{9}(O+N),
\end{equation}
where we have normalized the power spectrum by using for a flat
spectrum with a constant amplitude $B$:
\begin{equation}
[l(l+1)C_l]_{{\rm low}\,l}=\frac{9B}{100\pi}.
\end{equation}

We adopt the parameters
\begin{equation}
\label{cosmologparameters} \Omega_b\sim 0.04,\quad\Omega_\phi\sim
0.3,\quad\Omega_\Lambda\sim 0.7.
\end{equation}
We shall not attempt to separate the degeneracy of the parameter
$h$ in the Hubble expansion parameter: $H=100\, h\,{\rm km}/{\rm
Mpc}/{\rm sec}$ from the parameters $\Omega_b,\Omega_\phi$ and
$\Omega_\Lambda$, for this requires very accurate CMB power
spectrum data, which is not available at this time. We adopt in
the following the value $h\sim 0.71$. Moreover, we shall not
attempt accurate estimates of the parameters $r_h$ and $r_p$. It
is important to emphasize that the positions and heights of the
peaks depend sensitively on the values of the cosmological
parameters $\Omega_b,\Omega_\phi$ and $\Omega_\Lambda$. Indeed, a
fit to the CMB WMAP, Archeops and Boomerang data requires that
$\Omega_b <\Omega_m\sim\Omega_\phi < \Omega_\Lambda$ and that the
universe is spatially flat with
$\Omega_b+\Omega_\phi+\Omega_\Lambda=1$~\cite{WMAP,Archeops,Boomerang,Perlmutter,Riess}.

The oscillating power spectrum is obtained from the formula:
\begin{equation}
\label{fluct}
C(l)=\frac{1}{l^{1/2}}[119.11(0.053((\ln(0.0096l)-0.78)^2-4.3)\exp(2.13\times
10^{-7}l^2)\cos(0.01l+\frac{\pi}{4})
$$ $$
+0.14(0.5+0.36\ln(0.0096l))^2\cos(0.021l+\frac{\pi}{4}))\exp(-8.26\times
10^{-7}l^2]
$$ $$
+\frac{0.25(\ln(0.0096l)-0.024l^{0.3}-2.6)^2\exp(-4.01\times
10^{-7}l^2)}{1+5.59\times 10^{-5}l^{1.4}}
$$ $$
+\frac{0.28(\ln(0.096l)-0.027l^{0.3}+1.7)^2\exp(-8.26\times
10^{-7}l^2)}{1+5.59\times 10^{-5}l^{1.4}}
$$ $$
+\frac{0.18((\ln(0.0096l)-0.011l^{0.55}+2.2)^2\exp(-8.26\times
10^{-7}l^2)}{1+3.55\times 10^{-6}l^2}.
\end{equation}

The fluctuation spectrum determined by the analytical formula,
Eq.(\ref{fluct}), is displayed in Fig. 1 for the choice of
cosmological parameters given in (\ref{cosmologparameters}).
\vskip 0.3 in
\begin{center}\includegraphics[width=3in,height=3in]{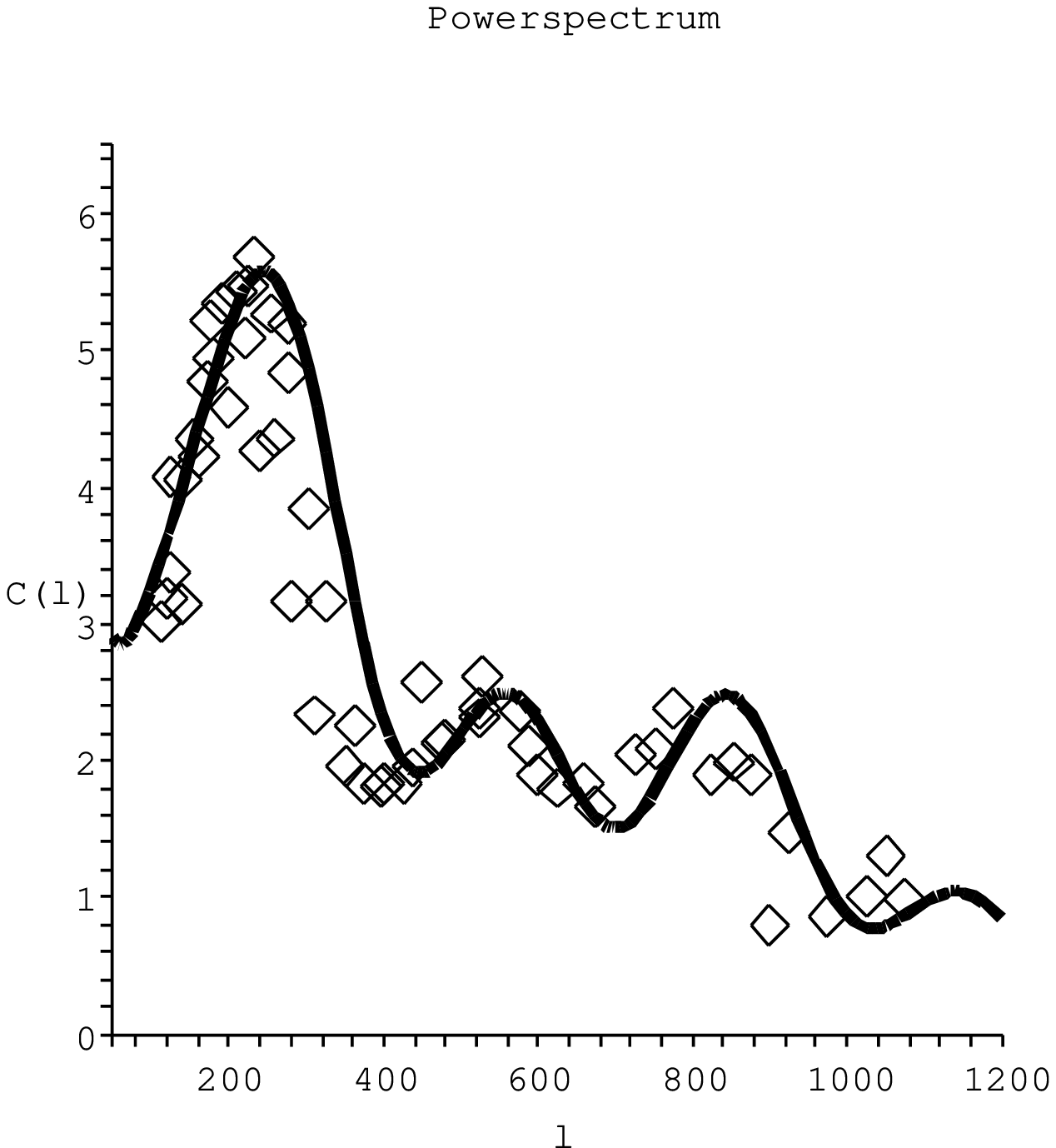}\end{center}
\vskip 0.1 in \begin{center} Figure 1. The solid line shows the
result of the calculation of the power spectrum acoustical
oscillations: $C(l)$, and the $\diamond s$ correspond to the WMAP,
Archeops and Boomerang data in units $\mu K^2\times
10^{-3}$~\cite{WMAP,Archeops,Boomerang}.
\end{center}
\vskip 0.1 in

Our predictions for the CMB power spectrum for large angular
scales corresponding to $l < 100$ will involve the integrated
Sachs-Wolfe contributions obtained from the gravitational
potential. These predictions should be calculated using a computer
code and will be investigated in a future article.
%\begin{equation}
%\label{bestparameters} \alpha_\infty=1.0\times 10^{-3},\quad
%\lambda_\infty=47\pm 1 AU,\quad {\bar r}=4.6\pm 0.2 AU,\quad
%b=4.0.
%\end{equation}
%\begin{figure}[h]
%\begin{center}
%\includegraphics{Cl4}%aulin, runningM, runningR, runningG, runningEta, yukawa
%\end{center}
%\caption{The analytical calculation of the power spectrum
%acoustical oscillations. The thick solid line displays the total
%acoustic oscillations result.The data are from the
%WMAP~\cite{WMAP} and Boomerang~\cite{Boomerang} observations.}
%\end{figure}
%\caption{\label{a_plog} Best fit to the Pioneer 10-11 anomalous
%acceleration data vs. the position, $\log{r}$ in AU.}
%\end{figure}
%\begin{figure}[h]
%\begin{center}
%\includegraphics{Bern/Paper/eps/aulin}%aulin, runningM, runningR, runningG, runningEta, yukawa
%\end{center}
%\caption{\label{a_p} Best fit to the Pioneer 10-11 anomalous
%acceleration data vs. the position, $r$ in AU.}
%\end{figure}

\section{Conclusions}

We have demonstrated that a modified gravity theory~\cite{Moffat2}
can lead to a satisfactory fit to the acoustical oscillation
spectrum obtained in the WMAP data~\cite{WMAP} by employing the
analytical formula for the fluctuation spectrum derived by
Mukhanov~\cite{Mukhanov}. The vector field $\phi^\mu$ (phion) and
the scalar fields $G$, $\omega$ and $\mu$ occur as new degrees of
freedom in the action of the MOG theory. The imperfect fluid
before recombination consists of two components, the baryon-photon
fluid and the neutral, non-relativistic phion BEC fluid. The
latter low-temperature, quantum state condensate fluid has zero
classical pressure and shear viscosity.

The analytical formula of Mukhanov~\cite{Mukhanov} reproduces
approximately the correct peak locations and peak heights in the
power spectrum data for the parameters: $\Omega_b\sim 0.04$,
$\Omega_m\sim\Omega_\phi\sim 0.3$, $\Omega_\Lambda\sim 0.7$ and
$H=71\,{\rm km}/{\rm Mpc}/{\rm sec}$. The varying constants $G$,
$\omega$ and $\mu=1/r_0=1/\lambda$ are assumed to have reached
their renormalized constant values before and shortly after
recombination. The analytical power spectrum fluctuations are
accurate enough to verify that our MOG gives a reasonable
description of the CMB spectrum for small angular values.

In the period before recombination and in cosmology at large
inter-galactic scales the quantum BEC phion energy density
dominates the matter density. The BEC fluid has zero classical
pressure and zero viscosity, whereby the phion field perturbations
as well as the scalar field perturbations are not subject to the
Silk dissipation that erases the baryon-photon perturbations. The
BEC fluid can describe a uniform density distribution that does
not couple to radiation and can become unstable due to the
attractive gravitational field and clump to form the seeds of
galaxies well before recombination, allowing enough time as the
universe expands to form clusters and super-clusters of galaxies.

An important problem to investigate is whether an N-body
simulation calculation based on our phion BEC scenario can predict
the observed large scale galaxy surveys. The formation of BEC
proto-galaxy structure before and after the epoch of recombination
and the growth of galaxies and clusters of galaxies at later times
in the expansion of the universe has to be explained. The BEC
condensates are not, perhaps, localizable as particles in the
sense of heavy WIMPS or axions.

For local late-time bound systems such as galaxies and clusters of
galaxies the symmetry breaking is relaxed and the phion
condensates can become ultra-light and relativistic. {\it Ordinary
baryonic matter and neutral hydrogen and helium gases} now
constitute the dominant form of matter. The extra degree of
freedom in MOG associated with the vector field $\phi^\mu$ and the
spatial variation of Newton's $G$ modifies for late-time bound
systems the Newtonian acceleration law for weak gravitational
fields. A calculation of the classical limit of the non-linear
Gross-Pitaevskii equation for the condensate wave function leads
to an effective, classical modified Newtonian acceleration law.
The rotational velocity curves of spiral galaxies are flattened,
because of the altered dynamics of the gravitational field at the
outer regions of the galaxies and not because of the presence of a
dominant dark matter halo.

The neutral phion BEC fluid dominates the cosmological density at
large cosmological scales, and the phion condensate in cosmology
has taken on the role of a light, cold ``dark matter'' fluid. This
allows for a fitting of the acoustical fluctuation spectrum at
small angular scales. According to our MOG scenario, we do not
expect in the present universe to detect heavy WIMPS such as the
supersymmetric neutralinos in accelerators or in underground
experiments. The dual role played by the phion particle in
describing galaxies and the large-scale structure of the universe
is a generic feature of our MOG theory.

We have succeeded in fitting in a unified picture a large amount
of data over 16 orders of magnitude in distance scale from Earth
to the surface of last scattering some 13.7 billion years ago,
using our modified gravitational theory. The data fitting ranges
over four distance scales: the solar system, galaxies, clusters of
galaxies and the CMB power spectrum data at the surface of last
scattering.

\vskip 0.2 true in {\bf Acknowledgments} \vskip 0.2 true in

This work was supported by the Natural Sciences and Engineering
Research Council of Canada. I thank Joel Brownstein, Clifford
Burgess, Martin Green and Constantinos Skordis for helpful
discussions.

\end{document}